\documentclass[1p,number,sort&compress]{elsarticle}

\usepackage{amsfonts}
\usepackage{amssymb}
\usepackage{amsmath}
\usepackage{graphicx}
\usepackage{moreverb}
\usepackage{url}
\usepackage{epsf}
\usepackage{color}
\usepackage{multirow}
\usepackage{subfigure}
\usepackage{bm}
\usepackage{pstricks}

\newcommand{\DZP}{$\mathrm{d}\zeta+\mathrm{p}$}
\newcommand{\QZTPDP}{$\mathrm{q}\zeta+\mathrm{tp}+\mathrm{dp}'$}
\newcommand{\SW}{\mathbf{S}_\mathrm{W}}
\newcommand{\HW}{\mathbf{H}_\mathrm{W}}
\newcommand{\bbx}{$\vcenter{\hbox{\includegraphics[width=10pt,height=10pt]{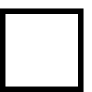}}}$}
\newcommand{\rct}{$\vcenter{\hbox{\includegraphics[width=10pt,height=10pt]{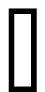}}}$}
\newcommand{\sbx}{$\vcenter{\hbox{\includegraphics[width=10pt,height=10pt]{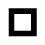}}}$}
\newcommand{\trr}{$\vcenter{\hbox{\includegraphics[width=10pt,height=10pt]{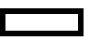}}}$}

\begin{document}
\title{The orbital minimization method for electronic structure calculations with finite-range atomic basis sets}
\author{Fabiano~Corsetti\corref{cor1}}
\cortext[cor1]{Corresponding author}
\ead{f.corsetti@nanogune.eu}
\address{CIC nanoGUNE, 20018 Donostia-San Sebasti\'{a}n, Spain}

\begin{abstract}
The implementation of the orbital minimization method (OMM) for solving the self-consistent Kohn-Sham (KS) problem for electronic structure calculations in a basis of non-orthogonal numerical atomic orbitals of finite-range is reported. We explore the possibilities for using the OMM as an exact cubic-scaling solver for the KS problem, and compare its performance with that of explicit diagonalization in realistic systems. We analyze the efficiency of the method depending on the choice of line search algorithm and on two free parameters, the scale of the kinetic energy preconditioning and the eigenspectrum shift. The results of several timing tests are then discussed, showing that the OMM can achieve a noticeable speedup with respect to diagonalization even for minimal basis sets for which the number of occupied eigenstates represents a significant fraction of the total basis size ($>15$\%). We investigate the hard and soft parallel scaling of the method on multiple cores, finding a performance equal to or better than diagonalization depending on the details of the OMM implementation. Finally, we discuss the possibility of making use of the natural sparsity of the operator matrices for this type of basis, leading to a method that scales linearly with basis size.
\end{abstract}


\maketitle

\section{Introduction}

Over the last two decades, density-functional theory~\cite{hk,ks} (DFT) has become a ubiquitous tool for studying molecular and condensed matter systems at the atomic level, with applications ranging from the Earth sciences to nanotechnology~\cite{intro1,intro2}. This is due in no small part to the proliferation of fast, accurate, easily available and user friendly software packages for performing DFT calculations (see, e.g., Refs.~\cite{Kresse1996a,Delley2000,Soler2002,castep,VandeVondele2005103,QE-2009,Blum20092175,abinit-generic}). Much work has gone into developing the methods used by such codes, and ongoing optimization of the underlying algorithms is essential to keep up with the possibilities offered by new computer architectures~\cite{Genovese2009,onetep-nick,Bowler2010,GPAW-GPU}.

Within the standard Kohn-Sham (KS) approach~\cite{ks}, the many-electron problem is reduced to a self-consistent eigenvalue problem with an effective Hamiltonian. When solving the problem in a basis, either repeated explicit diagonalizations (for small enough bases), or one of a number of iterative minimization algorithms~\cite{Wood1985,payne,Kresse1996a,Pfrommer1999287} can be used. The latter proceed either by minimizing the KS total energy functional directly, or as an alternative to diagonalization for a fixed Hamiltonian, within an outer self-consistency cycle for the electronic density. The use of an iterative algorithm of either form is essential for plane-wave methods, as the large number of basis functions/atom makes diagonalization prohibitively expensive for all but the smallest systems. One of the main advantages of using such algorithms is that only the occupied subspace needs to be computed, and for a plane-wave basis this typically corresponds to a very small fraction ($<1$\%) of the total number of eigenstates. Methods based on localized atomic-like orbitals~\cite{Delley2000,Soler2002,Blum20092175}, on the other hand, employ a much smaller number of basis functions/atom; in such cases, therefore, diagonalization is feasible even for large systems. Furthermore, the fraction of occupied eigenstates is much larger than for plane waves (10--20\%), making diagonalization not just competitive but indeed more efficient than most iterative algorithms.

In this paper, we present our implementation of an iterative minimization algorithm, which we refer to as the orbital minimization method (OMM) following Refs.~\cite{Tsuchida2007,Bowler2010}, as an alternative to explicit diagonalization in the SIESTA~\cite{Soler2002} code. The OMM works by finding the $N/2$ Wannier functions (WFs) describing the occupied subspace of an $N$-electron system by direct {\em unconstrained} minimization of an appropriately-constructed energy functional. This functional was originally proposed independently by Mauri, Galli and Car~\cite{Mauri1993,Mauri1994}, and Ordej\'{o}n {\em et al.}~\cite{Ordejon1993,Ordejon1995}, in the context of linear-scaling DFT methods~\cite{Bowler2010} with spatially confined WFs. In fact, a serial implementation of the linear-scaling OMM was an integral part of the original version of SIESTA; nowadays, however, most applications of the code employ diagonalization with LAPACK~\cite{laug}/ScaLAPACK~\cite{slug}. It is important to note that our new implementation is completely separate from this old one, and is {\em not} a linear-scaling solver; our aim, instead, is to explore the potential of the OMM as a conventional cubic-scaling iterative algorithm, which solves the KS problem exactly without introducing spatial truncations. Although previous investigations of cubic-scaling iterative algorithms in SIESTA (both Jacobi-Davidson~\cite{Sleijpen1996} and Lanczos-like~\cite{Lanczos1950}) have found that a very small fraction of occupied eigenstates is needed to rival the efficiency of diagonalization~\cite{BSC}, we show that this is indeed possible with the OMM even for a standard double-$\zeta$ basis with a single polarization shell (\DZP), for which the fraction of occupied eigenstates is significant, almost 20\%.

The use of the OMM as a cubic-scaling DFT solver has previously been described for a plane-wave basis by Pfrommer, Demmel and Simon~\cite{Pfrommer1999287} (alongside closely related methods), and implemented in the PARATEC~\cite{paratec} plane-wave code. In contrast, SIESTA makes use of a minimal basis of numerical atomic orbitals (NAOs) of finite range, leading to formally sparse Hamiltonian and overlap matrices. Our implementation is therefore somewhat different, in particular in the choice of preconditioner. Furthermore, the natural sparsity of the operators in our case can be used to eliminate the most expensive type of matrix--matrix multiplication present in the algorithm, leading to qualitatively different scaling behaviour with basis size.

The rest of the paper is organized as follows: in Sec.~\ref{sec:formal}, we describe in detail the implementation of the OMM and investigate its convergence properties. We give a theoretical overview of the method (Sec.~\ref{subsec:formal-over} and Sec.~\ref{subsec:formal-eta}), describe how the number of matrix operations can be minimized by careful consideration of the line search algorithm (Sec.~\ref{subsec:formal-ls}), discuss the issue of preconditioning (Sec.~\ref{subsec:formal-precon}) and empirically assess its efficiency (Sec.~\ref{subsec:formal-conv}), detail the use of sparse--dense matrix operations (Sec.~\ref{subsec:formal-sparse}), and comment on the issue of fractional occupancies (Sec.~\ref{subsec:formal-frac}). In Sec.~\ref{sec:timing}, we present scaling tests performed in serial and in parallel, and then investigate the hard and soft parallel scaling of the method (Sec.~\ref{subsec:timing-scaling}), and the scaling with basis size (Sec.~\ref{subsec:timing-basis}). Finally, in Sec.~\ref{sec:outro}, we give a summary of our main conclusions.

\section{Formalism and algorithmic considerations}
\label{sec:formal}

\subsection{OMM overview}
\label{subsec:formal-over}

We work in a basis of $m$ finite-range NAOs $\{ \phi_\mu (\mathbf{r}) \}$, typically (but not necessarily) atom-centred, and want to solve the generalized eigenvalue problem
\begin{equation} \label{eq:ge}
\mathbf{H} \mathbf{c}_\mu = \varepsilon_\mu \mathbf{S} \mathbf{c}_\mu,
\end{equation}
where
\begin{equation}
H_{\mu \nu} = \left \langle \phi_\mu \right | \hat{H}^\mathbf{KS} [\rho] \left | \phi_\nu \right \rangle,
\end{equation}
and
\begin{equation}
S_{\mu \nu} = \left \langle \phi_\mu | \phi_\nu \right \rangle.
\end{equation}
Within a single inner self-consistency (SCF) cycle, $\mathbf{H}$ depends on the fixed electronic density $\rho (\mathbf{r})$, and within an outer molecular dynamics (MD) step, both $\mathbf{H}$ and $\mathbf{S}$ depend on the atomic positions $\{ \mathbf{R}_I \}$.

Explicit diagonalization computes the KS eigenenergies $\{ \varepsilon_\mu \}$ and the matrix of KS eigenvectors $\mathbf{c}_\mu$, from which the full density matrix can be obtained. Instead, in the OMM we define a set of $n=N/2$ nonorthogonal WFs $\{ \chi_i (\mathbf{r}) \}$:
\begin{equation}
\left | \chi_i \right \rangle = \sum_{\mu=1}^m C^\mu_i \left | \phi_\mu \right \rangle.
\end{equation}
The reduced subspace operators defined by the WFs are then, in matrix form:
\begin{equation} \label{eq:HW}
\HW = \mathbf{C}^\mathrm{H} \mathbf{H} \mathbf{C}
\end{equation}
and
\begin{equation} \label{eq:SW}
\SW = \mathbf{C}^\mathrm{H} \mathbf{S} \mathbf{C}.
\end{equation}
We can also define a subspace energy $E$, which, when minimized with respect to the coefficients $\{ C^\mu_i \}$, will (by the variational principle) give the sum of the lowest $n$ eigenvalues of the original problem~\cite{Pfrommer1999287}:
\begin{equation} \label{eq:true_E}
E \left [ \mathbf{C} \right ] = 2\mathrm{Tr} \left \{ \SW^{-1} \HW \right \}
\end{equation}
(we include a factor of two for spin degeneracy).

The minimization of the functional given in Eq.~\ref{eq:true_E} has the advantage of being unconstrained, since no orthonormalization is required. Nevertheless, performing such a minimization is computationally demanding due to the presence of the inverse overlap $\mathbf{S}^{-1}_\mathrm{W}$, which needs to be recomputed at every trial step; furthermore, this means that the line search in a steepest descent (SD) or conjugate gradient (CG) algorithm has to be solved numerically.

The OMM substitutes Eq.~\ref{eq:true_E} with a different functional, one that does not contain the inverse operation:
\begin{equation} \label{eq:OMM_E}
\tilde{E} \left [ \mathbf{C} \right ] = 2\mathrm{Tr} \left \{ \left [ \mathbf{I}_n + \left ( \mathbf{I}_n-\SW \right ) \right ] \HW \right \} = 4\mathrm{Tr} \left \{ \HW \right \} -2\mathrm{Tr} \left \{ \SW \HW \right \}.
\end{equation}
It can be shown that this new functional drives the WFs towards orthonormality as it is minimized~\cite{Mauri1993,Mauri1994,Ordejon1993,Ordejon1995}; at the minimum, therefore, $\SW=\mathbf{I}_n$, and $\tilde{E} \left [ \mathbf{C}_0 \right ] = E \left [ \mathbf{C}_0 \right ] = E_0$, where $\mathbf{C}_0$ describes the occupied subspace, and $E_0$ is the ground-state KS (band) energy. Eq.~\ref{eq:OMM_E} can be derived either by replacing the inverse overlap matrix in Eq.~\ref{eq:true_E} with a first-order Taylor expansion~\cite{Mauri1993}, or by using a Lagrange multiplier approach to enforce the desired orthonormality requirement on $\mathrm{Tr} \left \{ \HW \right \}$ at the solution~\cite{Ordejon1993}.

The OMM functional, therefore, allows for unconstrained minimization without requiring any matrix inversion; this is particularly suitable for developing linear-scaling methods, since the inverse of a formally sparse $\SW$ matrix (obtained by constraining the radii of the WFs) will not itself be sparse. However, the OMM in its original form was quickly abandoned by the linear-scaling community, as the localization of the WFs introduces many local minima that were found to lead to serious difficulties in obtaining the true ground state~\cite{Mauri1994,Ordejon1995,Kim1995}. Ultimately, this led to the development and implementation of various generalized OMMs to overcome the convergence problem~\cite{Hierse1994,Kim1995,Yang1997,Tsuchida2007}. For non-linear-scaling implementations of the OMM, however, this problem does not present itself, as the WFs are not spatially constrained; the simplicity of the original functional is therefore ideal in our case for developing an efficient algorithm.

\subsection{Eigenspectrum shift}
\label{subsec:formal-eta}

As shown by Ordej\'{o}n {\em et al.}~\cite{Ordejon1995} and Kim, Mauri and Galli~\cite{Kim1995}, a stationary point of the OMM functional is obtained when all the WFs are either eigenvectors of Eq.~\ref{eq:ge} or zero, subject to any arbitrary unitary transformation between them. If all the corresponding eigenvalues are negative, this point will be a minimum; however, if any of them are positive, it will become a saddle point. Therefore, the stationary point at $\mathbf{C}_0$ will be a minimum provided that all the occupied eigenvalues are negative\footnote{Furthermore, it can be shown that it will only be a global minimum if the entire eigenspectrum of $\hat{H}$ is negative; otherwise, it will be a local minimum, and the functional will have no lower bound.}.

Because of this, it is necessary to shift the eigenspectrum by $\eta > \varepsilon_n$ for the minimization procedure to be able to find the correct ground state. This is achieved by the transformation
\begin{equation}
\mathbf{H} \rightarrow \mathbf{H}-\eta\mathbf{S},
\end{equation}
and the corresponding modification to the functional
\begin{equation}
\tilde{E} \rightarrow \tilde{E}+n\eta.
\end{equation}
Pfrommer {\em et al.} have analyzed the efficiency of the OMM convergence as a function of $\eta$, showing that the optimal choice (for minimizing the condition number of the associated Hessian matrix) lies within the range
\begin{equation} \label{eq:eta}
\frac{\varepsilon_{n+1}-\varepsilon_n}{4}+\varepsilon_n \le \eta \le \frac{\varepsilon_m-\varepsilon_1}{4}+\varepsilon_1;
\end{equation}
therefore, if the width of the occupied bands is $\gtrsim 25$\% of that of the total eigenspectrum, it will not be possible to find a value of $\eta$ satisfying these conditions. This will generally only be the case for a truly minimal (single-$\zeta$) basis; even in such cases, however, we have not observed in practice any noticeable decrease in efficiency (see Sec.~\ref{subsec:formal-conv-opt} for further discussion).

\subsection{Line search}
\label{subsec:formal-ls}

For the minimization of the OMM functional we use a CG scheme, with the conjugate search direction $\mathbf{D}$ given by the Polak-Ribi\'{e}re formula~\cite{Polak1971}. The gradient of the OMM functional used for this scheme can be calculated as:
\begin{equation} \label{eq:G}
\mathbf{G} = 8 \mathbf{H} \mathbf{C} -4 \mathbf{S}\mathbf{C}\HW -4 \mathbf{H}\mathbf{C}\SW.
\end{equation}

A particularly convenient property of the functional is that it is a quartic function along a given search direction. This means that the line search can be solved exactly, either by fitting to five energy/gradient points along the line, or by directly computing the coefficients $\left \{ \alpha_0, \ldots, \alpha_4 \right \}$ of the fourth-order polynomial. For the former approach, previous implementations have used five energy points~\cite{Ordejon1993}, or four energy points and the initial gradient point~\cite{Ordejon1995}. For the latter approach, which is employed in our current implementation, we give here the simplified expressions for calculating the coefficients:

\begin{equation}
\begin{cases}
\alpha_0 =& \tilde{E} \left [ \mathbf{C} \right ] \\
\alpha_1 =&  8\mathrm{Tr} \left \{ \HW' \right \} -4\mathrm{Tr} \left \{ \SW'\HW \right \} -4\mathrm{Tr} \left \{ \SW\HW' \right \} \\
\alpha_2 =&  4\mathrm{Tr} \left \{ \HW'' \right \} -2\mathrm{Tr} \left \{ \SW''\HW \right \} -2\mathrm{Tr} \left \{ \SW\HW'' \right \} \\
          & -4\mathrm{Tr} \left \{ \SW'\HW' \right \} -4\mathrm{Tr} \left \{ \left ( \SW' \right )^\mathrm{H}\HW' \right \} \\
\alpha_3 =& -4\mathrm{Tr} \left \{ \SW''\HW' \right \} -4\mathrm{Tr} \left \{ \SW'\HW'' \right \} \\
\alpha_4 =& -2\mathrm{Tr} \left \{ \SW''\HW'' \right \}
\end{cases},
\end{equation}
where
\begin{equation}
\HW' = \mathbf{D}^\mathrm{H} \mathbf{H} \mathbf{C}
\end{equation}
and
\begin{equation}
\HW'' = \mathbf{D}^\mathrm{H} \mathbf{H} \mathbf{D},
\end{equation}
and similarly for $\SW'$ and $\SW''$ (in this case, $\mathbf{D}$ is the line search direction and $\mathbf{C}$ is the starting point). Once the position of the line minimum $x$ is solved for~\cite{nr}, the new starting point for the following line search is easily computed:
\begin{equation}
\begin{cases}
\mathbf{C}^\mathrm{new} = \mathbf{C} + x\mathbf{D} \\
\HW^\mathrm{new} = \HW + x\HW' + x\left ( \HW' \right )^\mathrm{H} + x^2\HW''
\end{cases},
\end{equation}
and similarly for $\SW^\mathrm{new}$.

\begin{table*}
\begin{tabular*}{\textwidth}{lcccccccc}
\hline
\hline
\\[-10pt]
                              & \multicolumn{2}{c}{Computational cost} & Fit $5+0$                 & Fit $4+1$                 & Fit       &            \\ \cline{2-3}
                    Operation & Num. $\times$ & Num. $+$               & (Ref.~\cite{Ordejon1993}) & (Ref.~\cite{Ordejon1995}) & $3+2$     & Coeffs.    \\
\\[-10pt]
\hline \\[-10pt]
              \bbx\rct$=$\rct &        $fm^3$ &            $fm^3-fm^2$ & $\times$10                & $\times$8                 & $\times$6 &  $\times$2 \\
              \rct\sbx$=$\rct &      $f^2m^3$ &          $f^2m^3-fm^2$ &  $\times$2                & $\times$2                 & $\times$4 &  $\times$2 \\
              \trr\rct$=$\sbx &      $f^2m^3$ &        $f^2m^3-f^2m^2$ & $\times$10                & $\times$8                 & $\times$6 &  $\times$4 \\
   \rct$+\ \alpha$\rct$=$\rct &        $fm^2$ &                 $fm^2$ &  $\times$6                & $\times$5                 & $\times$4 &  $\times$3 \\
vec\{\rct\}$\cdot$vec\{\rct\} &        $fm^2$ &               $fm^2-1$ &          -                & $\times$1                 & $\times$2 &          - \\
   \sbx$+\ \alpha$\sbx$=$\sbx &      $f^2m^2$ &               $f^2m^2$ &          -                &         -                 &         - &  $\times$6 \\
               Tr\{\sbx\sbx\} &      $f^2m^2$ &             $f^2m^2-1$ &  $\times$5                & $\times$4                 & $\times$3 & $\times$10 \\
                   Tr\{\sbx\} &             - &                 $fm-1$ &  $\times$5                & $\times$4                 & $\times$3 &  $\times$3 \\
\\[-10pt]
\hline
\hline
\end{tabular*}
\caption{Number of matrix operations needed for a CG line search, for three different fitting strategies and for direct computation of the coefficients of the quartic function. `Fit $a+b$' refers to a fit of $a$ energy points and $b$ gradient points. Listed are the type of operation, the number of elementary multiplications and additions it requires, and the number of times it is used in a particular scheme. The matrices shown schematically are: \protect\bbx\ $\mathbf{H}$, $\mathbf{S}$ ($m \times m$); \protect\sbx $\HW$, $\SW$ ($fm \times fm$); \protect\rct $\mathbf{C}$, $\mathbf{D}$ ($m \times fm$). $f=n/m$ is the fraction of occupied eigenstates. The cost of calculating the conjugate direction from the initial gradient is not included, and is the same for all methods. We note that we do not make use of the Hermiticity of the operators.}
\label{table:ops}
\end{table*}

Table~\ref{table:ops} lists the number of matrix operations needed for solving the line search with these different approaches. The three types of matrix--matrix multiplication listed are the only cubic-scaling operations; of these, the one involving an $m \times m$ operator matrix in the NAO basis ($\mathbf{H}$ or $\mathbf{S}$) is the most expensive, generally by an order of magnitude. When using the fitting approach, therefore, the cost of the line search is minimized by choosing three energy points and two gradient points, since calculating the gradient does not require this operation if the energy has already been calculated at the same point\footnote{In this discussion we assume that the results of the intermediate multiplications $\left ( \mathbf{H}\mathbf{C} \right )$, $\left ( \mathbf{S}\mathbf{C} \right )$, $\left ( \mathbf{H}\mathbf{D} \right )$, and $\left ( \mathbf{S}\mathbf{D} \right )$ are saved and then reused for calculating $\mathbf{G}$, $\HW'$, $\SW'$, $\left ( \mathbf{H}\mathbf{C} \right )^\mathrm{new} = \left ( \mathbf{H}\mathbf{C} \right ) + x\left ( \mathbf{H}\mathbf{D} \right )$, and $\left ( \mathbf{S}\mathbf{C} \right )^\mathrm{new} = \left ( \mathbf{S}\mathbf{C} \right ) + x\left ( \mathbf{S}\mathbf{D} \right )$.}.

Calculating the coefficients of the quartic function directly has two advantages over fitting to data points sampled along the line: firstly, it eliminates the necessity of choosing a step length, which can pose some dangers from the numerical point of view (i.e., too large or too small a sampling range can result in a significant error in the computed value of $x$, sometimes failing entirely to find an existing minimum); secondly, and most importantly, it reduces the number of operations that need to be performed almost by a factor of three. We note, however, that we do not expect this approach to be efficient for linear-scaling implementations of the OMM, because the matrix sparsity will be progressively reduced from $\HW$ to $\HW'$ to $\HW''$, and all non-zero elements of the latter two matrices are needed for calculating the coefficients.

\subsection{Preconditioning}
\label{subsec:formal-precon}

The problem of kinetic energy ill-conditioning is well known for plane-wave electronic structure calculations~\cite{payne}; however, it is also present for localized orbitals, as the high-energy unoccupied eigenstates are similarly dominated by the kinetic energy contribution. A preconditioner suitable for localized orbitals was proposed by Bowler and Gillan~\cite{Bowler1998}, and then modified by Gan, Haynes and Payne~\cite{Gan2001} to respect the tensorial nature of the search direction (which becomes apparent when using a non-orthogonal basis~\cite{White1997}). Following the latter, we premultiply the gradient $\mathbf{G}$ by the preconditioning matrix:
\begin{equation}
\mathbf{P} = \left ( \mathbf{S}+\frac{1}{\tau}\mathbf{T} \right )^{-1},
\end{equation}
where $\mathbf{T}$ is the kinetic energy matrix in the NAO basis, and $\tau$ sets the scale for the kinetic energy preconditioning. For $\tau = \infty$, the preconditioner becomes simply the inverse of the overlap matrix, and, therefore, acts as a pure tensorial correction transforming the covariant gradient defined in Eq.~\ref{eq:G} into the contravariant gradient suitable for updating the coefficients $\mathbf{C}$. Within a similar scheme, Mostofi {\em et al.}~\cite{Mostofi2003} also postmultiply $\mathbf{PG}$ by $\SW$ to account for the non-orthogonality of the WF basis in addition to that of the underlying basis; however, this is not possible for the OMM, as the functional itself does not respect tensorial rules for the WFs.

From the point of view of computational expense, the matrix inverse needed to find $\mathbf{P}$ does not contribute significantly to the total, since it is only performed once at the beginning of each MD step. However, calculating the preconditioned gradient adds an operation of type \bbx\rct$=$\rct (Table~\ref{table:ops}) for each line search. This added expense is usually more than compensated by the decrease in the number of line searches, as we will show. An alternative approach is to reduce the generalized eigenvalue problem to standard form by Cholesky factorization. This is the preferred method for cases in which kinetic energy preconditioning is not needed, as it can be shown that the minimization of the reduced Hamiltonian exactly follows that of the original problem with the application of the $\mathbf{S}^{-1}$ preconditioner; computationally, however, it is cheaper, since (a) no premultiplication of the gradient is necessary, and (b) the calculation of $\SW$ is reduced to $\mathbf{C}^\mathrm{H} \mathbf{C}$, thus halving the number of \bbx\rct$=$\rct operations listed in Table~\ref{table:ops}.

\subsection{Convergence tests}
\label{subsec:formal-conv}

\subsubsection{Optimal values of $\tau$ and $\eta$}
\label{subsec:formal-conv-opt}

\begin{figure}[t]
\begin{center}\includegraphics{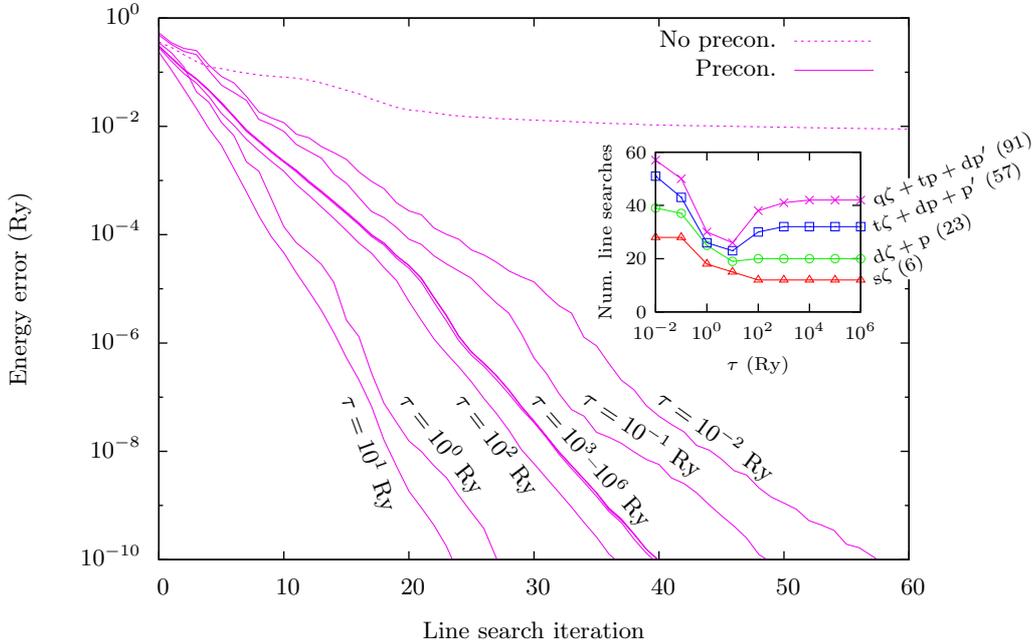}\end{center}
\caption{Convergence of the error in the OMM energy for the water molecule with respect to the ground-state band energy calculated by explicit diagonalization (only a single SCF iteration is considered). The inset shows the number of line search iterations needed to reach a fixed convergence threshold, as a function of $\tau$ for four different basis sets (the number of basis orbitals is given in brackets). The main panel shows the results for the \QZTPDP\ basis. The basis set notation is explained in the text.}
\label{fig:tau}
\end{figure}

\begin{figure}[t]
\begin{center}\includegraphics{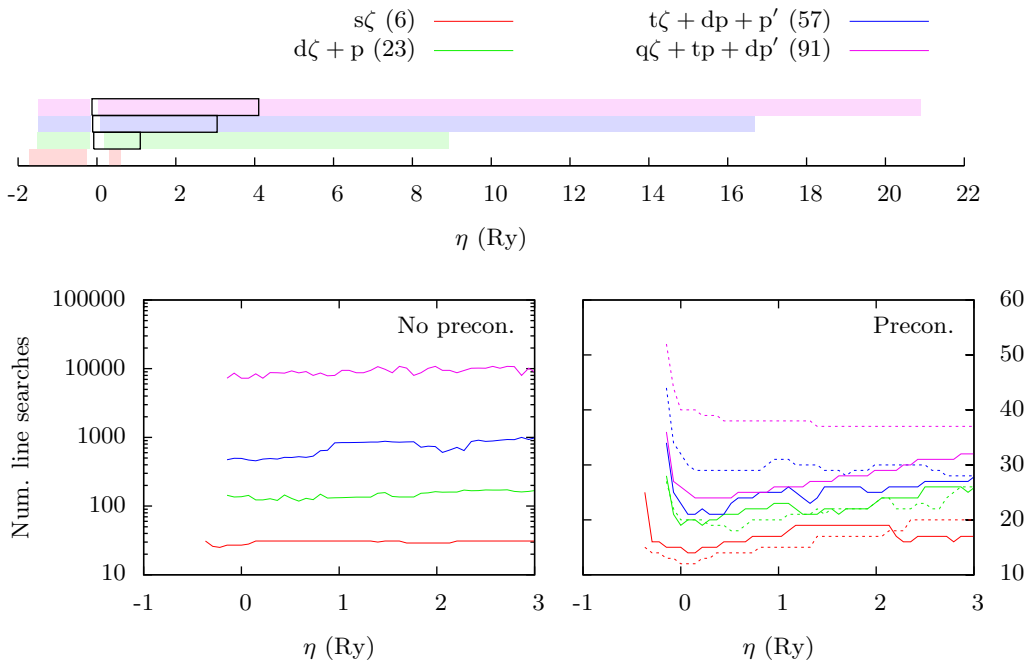}\end{center}
\caption{Number of line search iterations needed to reach a fixed convergence threshold for the water molecule, as a function of the eigenvalue shift parameter $\eta$ for four different basis set. The upper panel shows the width of the occupied and unoccupied portions of the eigenspectrum for each basis (the light coloured boxes), and the range of optimal $\eta$ satisfying Eq.~\ref{eq:eta} (the empty black box). The number of basis orbitals is given in brackets in the key. The basis set notation is explained in the text. The lower left panel shows the results of the minimization without preconditioning; the lower right panel shows the results with preconditioning, for $\tau = 10$~Ry (solid lines) and $\tau = \infty$ (dashed lines).}
\label{fig:eta}
\end{figure}

Fig.~\ref{fig:tau} shows a simple test case for our implementation of the OMM in SIESTA: the convergence of the energy $\tilde{E} [\mathbf{C}]$ for a water molecule in a cubic box of side $10$~\AA. We show the convergence of the CG minimization algorithm without preconditioning, and with preconditioning for different values of $\tau$. We have repeated the calculations for four different basis sets, ranging from the smallest single-$\zeta$ basis to a large and extremely precise quadruple-$\zeta$ basis with two shells of polarization orbitals\footnote{We employ the following notation for naming the bases: $\mathrm{A}\zeta+\mathrm{Bp}+\mathrm{Cp}'$, where A indicates the number of $\zeta$ orbitals used for the valence shells (s: single, d: double, t: triple, q: quadruple), B for the first polarization shell (if present), and C for the second one (if present). For single-$\zeta$ polarization shells, the s prefix is omitted. For more information on the bases used for the water molecule, see Ref.~\cite{basis_us}.}. The fraction of occupied eigenstates ranges from 67\% for the former basis, down to 4\% for the latter. The occupied and unoccupied portions of the total eigenspectrum for the four bases are shown in the upper panel of Fig.~\ref{fig:eta}.

The main panel of Fig.~\ref{fig:tau} shows the energy convergence for the largest basis. In this case, the preconditioner is essential for achieving convergence within a reasonable number of line search iterations. This is due almost entirely to the tensorial correction, as shown by the convergence behaviour in the limit of large $\tau$. Nevertheless, the choice of $\tau$ provides some additional speedup, with the best performance given by $\tau \sim 1$--10~Ry (consistent with the idea of setting $\tau$ equal to the highest kinetic energy of the occupied eigenstates~\cite{Gan2001}, in this case 2.85~Ry).

As should be expected, the maximum speedup provided by kinetic energy preconditioning increases with the size of the basis, while the optimal value of $\tau$ is system-dependent and so does not change noticeably between bases (see inset of Fig.~\ref{fig:tau}). For very small (single- and double-$\zeta$) bases, it provides little or even no benefit, but it becomes increasingly important when using larger triple- and quadruple-$\zeta$ bases. In fact, while the number of iterations needed to reduce the energy error to within a given tolerance increases steadily with basis size when using the pure tensorial correction, it appears instead to converge towards a fixed number when kinetic energy preconditioning is also included.

We now return to the question of the optimal choice of the eigenvalue shift parameter, $\eta$ (Sec.~\ref{subsec:formal-eta}), and its interaction with the preconditioner. Fig.~\ref{fig:eta} shows the number of line search iterations needed for convergence of the water molecule using the four basis sets described previously, both without and with preconditioning (lower left and right panels, respectively). The eigenspectrum and the corresponding `optimal' range of $\eta$ satisfying Eq.~\ref{eq:eta} are shown for each basis in the upper panel. For the single-$\zeta$ basis, no $\eta$ can be found to satisfy these conditions, due to the narrow width of the unoccupied eigenspectrum.

For non-preconditioned minimization, the choice of $\eta$ has almost no visible effect in the logarithmic plot shown in Fig.~\ref{fig:eta}; in fact, the ill-conditioning due to the tensorial incorrectness of the gradient is much more important. Even so, in general the number of line searches required for convergence is seen to increase as $\eta$ is raised above $\sim$$\varepsilon_{n+1}$ (the lowest unoccupied eigenstate energy), with all bases exhibiting approximately similar behaviour.

\begin{figure}
\begin{center}\subfigure[Basis size variation (64-atom cell of bulk Si, num. of basis orbitals/atom given in brackets)]{\includegraphics{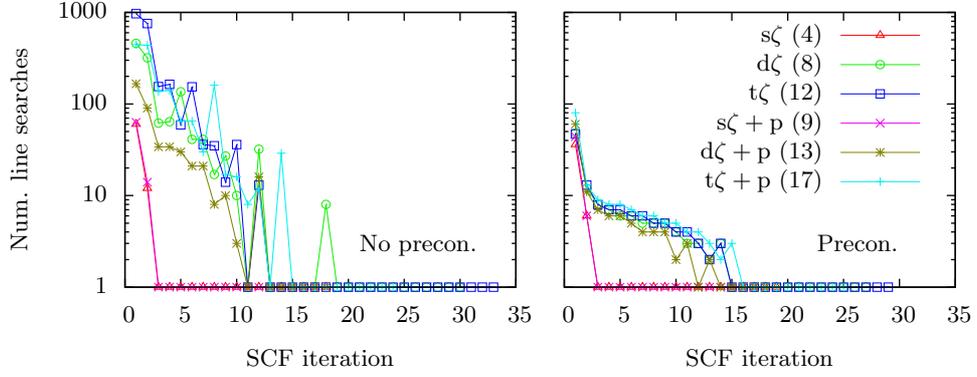}}
\subfigure[System size variation (bulk Si with \DZP\ basis)]{\includegraphics{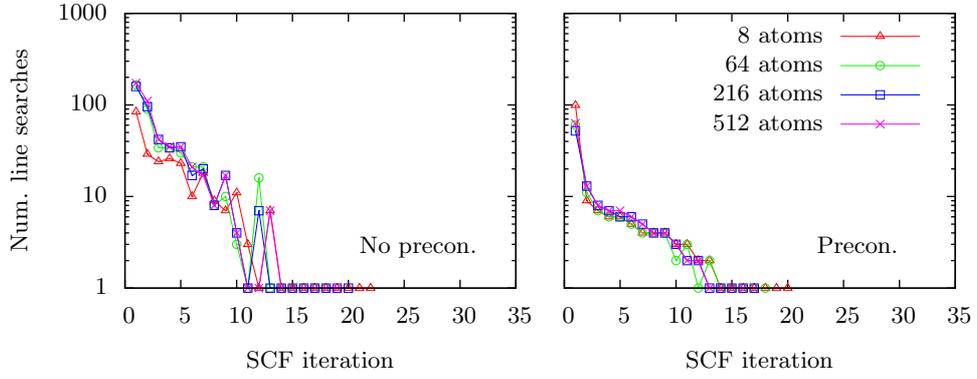}}
\subfigure[Material variation (512-atom cell of bulk material with \DZP\ basis, band gap given in brackets)]{\includegraphics{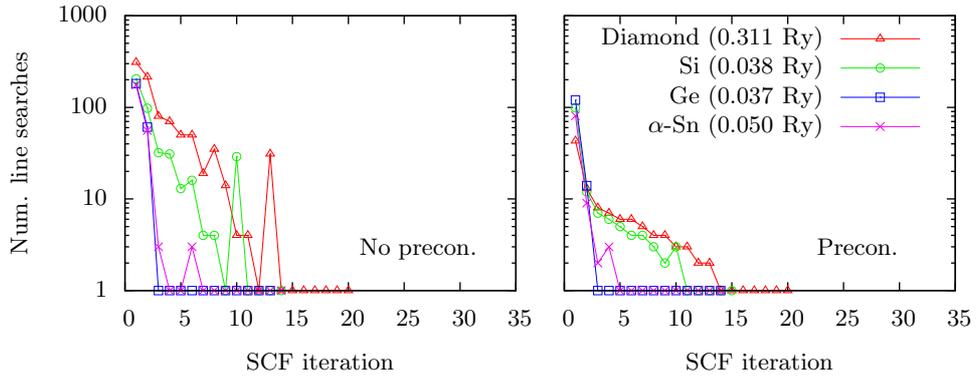}}\end{center}
\caption{Convergence of the OMM without (left panel) and with (right panel) preconditioning for a variety of different systems. $\tau=10$~Ry for all preconditioned simulations.}
\label{fig:precon}
\end{figure}

The case of the preconditioned CG algorithm is more interesting: the pure tensorial correction ($\tau = \infty$) removes the major source of ill-conditioning while not affecting the width of the eigenspectrum, in doing so revealing a behaviour that confirms the analysis of Pfrommer {\em et al.}; for the smaller bases the number of line searches increases steadily as $\eta$ is raised (and it does so fastest for the smallest basis), while for the larger bases there is no increase, since the range of $\eta$ shown in the figure lies within the optimal range given by Eq.~\ref{eq:eta}.

The application of kinetic energy preconditioning ($\tau = 10$~Ry in this example) has the effect of compressing the unoccupied eigenspectrum to approximately the same size for all bases; the result, as should be expected, is that the effect of varying $\eta$ is also almost independent of basis. This behaviour is readily apparent in the figure. Unfortunately, as reducing the width of the eigenspectrum necessarily reduces the optimal range of $\eta$, this means that the choice of $\eta$ becomes important even for large bases. The best choice is found around the bottom of the unoccupied eigenspectrum. However, the increase in the number of iterations with $\eta$ is quite slow (2--3 additional line searches per Ry increase of $\eta$); as the Fermi level in SIESTA does not vary by more than 1~Ry for any reasonable physical system, this should not pose a problem in practical applications of the OMM.

\subsubsection{Self-consistent calculations}
\label{subsec:formal-conv-SCF}

Finally, we discuss the performance of the OMM for realistic self-consistent single-point energy calculations, typically requiring $\sim$20 SCF cycles. In common with other iterative minimization algorithms, the OMM can reuse the final set of coefficients $\mathbf{C}$ from one SCF step as the starting guess for the next one, progressively reducing the number of line search iterations needed for each cycle. As shown in the examples in Fig.~\ref{fig:precon}, the last 5--10 SCF cycles in a self-consistent calculation require a single OMM line search. Our stopping criterion for the minimization procedure is given by the relative energy difference between subsequent line searches:
\begin{equation}
2\frac{\tilde{E} \left [ \mathbf{C}^\mathrm{new} \right ]-\tilde{E} \left [ \mathbf{C} \right ]}{\tilde{E} \left [ \mathbf{C}^\mathrm{new} \right ]+\tilde{E} \left [ \mathbf{C} \right ]} \le e^\mathrm{tol};
\end{equation}
we use a convergence threshold of $e^\mathrm{tol}=10^{-9}$. The total interacting energy of the self-consistent system for all the examples shown is in close agreement with that obtained by explicit diagonalization (with discrepancies of $< 10^{-6}$~Ry/atom, on the same order as the tolerance in the SCF convergence).

Fig.~\ref{fig:precon} shows the difference in convergence between non-preconditioned and preconditioned minimization when varying either (a) the basis size, (b) the system size, or (c) the material. The tests are performed on bulk crystalline silicon, except for (c), in which other group IV elements with the same diamond crystal structure are also used. $\eta$ is set to zero for all systems, except for germanium, for which it is set to 0.5~Ry. When using the preconditioner, $\tau$ is set to 10~Ry for all systems.

The preconditioner is extremely effective, not only in reducing the number of line searches, but also in stabilizing the convergence. This is most important for curing the ill-conditioning caused by large basis sets, as discussed previously, but also has a noticeable effect when varying the material. The system size variation, instead, is already quite stable, with the non-preconditioned minimization giving almost identical convergence behaviour for all systems with $\ge 64$ atoms.

\subsection{Sparse algebra}
\label{subsec:formal-sparse}

Although we are not developing a linear-scaling solver, we may still take advantage of the formal sparsity of the $\mathbf{H}$ and $\mathbf{S}$ matrices in SIESTA to reduce the computational expense of the algorithm. In fact, in the description of the line search method given in Sec.~\ref{subsec:formal-ls}, there are four operations of type \bbx\rct$=$\rct that are multiplications of an $m \times m$ sparse matrix with an $m \times n$ dense matrix ($\mathbf{H} \mathbf{C}$, $\mathbf{H} \mathbf{D}$, $\mathbf{S} \mathbf{C}$, $\mathbf{S} \mathbf{D}$). We can therefore substitute the dense--dense multiplications with sparse--dense ones, reducing the cost of these operations from $\mathcal{O} \left ( m^3 \right )$ to $\mathcal{O} \left ( m^2 \right )$. The overall scaling of the solver, however, is still $\mathcal{O} \left ( m^3 \right )$; hence, the details of the sparse--dense multiplication algorithm (which we have developed internally to SIESTA to conform to its native sparse matrix representation~\cite{Soler2002}) are not particularly important, as this operation will no longer constitute the computational bottleneck other than for small systems, for which entirely dense multiplication is preferable.

The $\mathbf{P}$ matrix, also $m \times m$, is not formally sparse; attempts to truncate it (e.g., by setting all elements below a threshold to zero) have not been successful in retaining the advantage of preconditioning. We therefore leave $\mathbf{P} \mathbf{G}$ as the only dense--dense multiplication of its kind. Similarly, Cholesky factorization tends to destroy the sparsity of the reduced $\mathbf{H}$, and so no sparse--dense operations remain at all in this approach\footnote{We note that sparsity-preserving factorizations have been developed~\cite{George1988,Chen2008}, although we do not explore them in the current study.}.

\subsection{Fractional occupancies}
\label{subsec:formal-frac}

The issue of fractional occupancies is generally connected to that of finite-temperature (Fermi-level smearing) calculations, in particular for achieving numerical stability of the SCF convergence in metallic systems~\cite{Kresse199615,Springborg1998}. A given smearing function can easily be applied with a knowledge of the KS eigenenergies around the Fermi level, and their corresponding eigenvectors; however, methods that do not explicitly make use of this information have also been proposed, especially in the context of linear-scaling applications~\cite{Goedecker1994,Goedecker1995,Corkill1996}.

The current implementation of the OMM as described in this paper does not support Fermi-level smearing, since the method does not natively provide information on individual eigenstates. We shall not discuss this issue further here, but note that we plan to address it in future; a possible simple method to do so would be as follows: (i) isolate the Hilbert subspace for a small number of eigenstates above and below the Fermi level, by two nested minimizations; (ii) solve for individual eigenstates in this subspace (either by explicit diagonalization or band-by-band minimization) and apply the appropriate fractional occupations to them.

Finally, it is important to note that the current implementation is nevertheless robust for Hamiltonians with a degenerate ground state: tests have shown no difference in the rate of convergence, and the correct band energy being recovered. However, the degenerate eigenstates at the Fermi level will be partially filled in a random (i.e., non-thermodynamic) combination, thereby resulting in a lowering of symmetry of the output charge density. While this leads to instabilities in the SCF convergence for metallic systems, we can expect it to be less problematic for the case of accidental degeneracies appearing during the course of the self-consistency cycle.

\section{Timing tests}
\label{sec:timing}

\begin{figure}[t]
\begin{center}\includegraphics{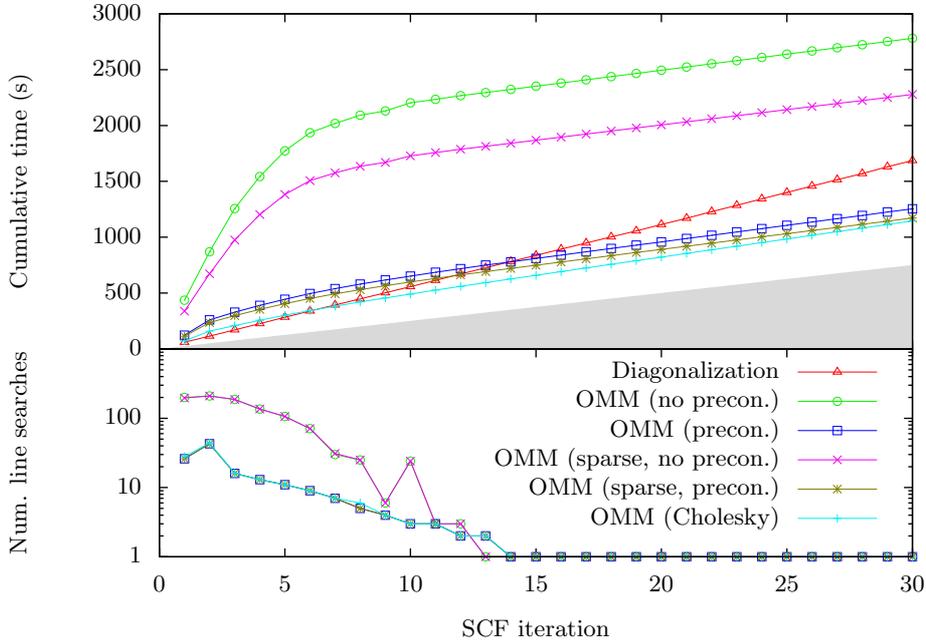}\end{center}
\caption{Timing test for a single-point energy calculation of 200 molecules of liquid water. The cumulative time at each SCF iteration is shown in the upper panel, and the number of OMM line searches performed for each step is shown in the lower panel. The shaded gray area shows the cumulative time spent on operations outside of the solver routine (diagonalization/OMM), that are the same for all calculations. $\tau=10$~Ry for all preconditioned simulations.}
\label{fig:single_step}
\end{figure}

In this section, we present some representative results obtained for the OMM, comparing its performance in each case to that of explicit diagonalization using (Sca)LAPACK (see Sec.~\ref{subsec:timing-scaling} for more details). We consider five different variants of the OMM algorithm, as listed in the key of Fig.~\ref{fig:single_step}:
\begin{itemize}
\item OMM (no precon.): entirely dense matrix multiplications using (P)BLAS~\cite{blas,slug}, without preconditioning;
\item OMM (precon.): entirely dense matrix multiplications using (P)BLAS, with preconditioning (matrix inverse using (Sca)LAPACK);
\item OMM (sparse, no precon.): \bbx\rct$=$\rct operations performed sparse--dense, others entirely dense using (P)BLAS, without preconditioning;
\item OMM (sparse, precon.): \bbx\rct$=$\rct operations performed sparse--dense except the preconditioning operation $\mathbf{PG}$, performed dense--dense;
\item OMM (Cholesky): entirely dense matrix multiplications using (P)BLAS, Cholesky factorization using (Sca)LAPACK.
\end{itemize}

For all our test systems, we use norm-conserving Troullier-Martins pseudopotentials~\cite{Troullier1991} in separable Kleinman-Bylander form~\cite{pseudo-KB} (with a maximum angular momentum component of $l=3$), and, unless otherwise stated, \DZP\ basis sets. We use the LDA~\cite{qmc2} exchange-correlation functional, and represent the electronic density on a real-space grid with a grid cutoff~\cite{Soler2002} of 100~Ry, except for our simulations of liquid water, for which we use the non-local vdW-DF functional of Dion {\em et al.}~\cite{vdW-DF}, and a grid cutoff of 150~Ry. The simulation cell is periodic, and, unless otherwise stated, only the $\Gamma$-point is used.

Fig.~\ref{fig:single_step} shows the timing results for a single-point energy calculation of a snapshot of 200 molecules of liquid water, extracted from an {\em ab initio} molecular dynamics (AIMD) simulation. The matrix dimensions are $m=4600$ and $n=800$, and the level of sparsity of the $\mathbf{H}$ and $\mathbf{S}$ matrices is approximately 94\%. The test was run on 10 cores of a Dell PowerEdge R910 server with four Intel Xeon E7-4850 processors. The code was compiled using Open MPI and OpenBLAS~\cite{openblas}.

The simulation takes 30 SCF iterations to reach self-consistency, within SIESTA's default tolerance on the density matrix; the plot shows the cumulative time at each step. As should be expected, diagonalization uses a fixed amount of time per SCF step, leading to a linear increase in time against iteration number. The OMM, instead, benefits from the reuse of information, and so the time taken per step decreases steadily during the first $\sim$10--15 steps, until reaching a constant minimum value, corresponding to a single line search per SCF iteration. These single-line search steps are $\sim$4--14 times faster for the OMM than for diagonalization depending on the specific algorithm used, as shown in Table~\ref{table:single_step}. The performance of the different OMM flavours reflects the number of dense--dense \bbx\rct$=$\rct operations needed per line search, as discussed previously. Cholesky factorization is the most expensive method for performing a single line search, due to the extra cost of transforming the $\mathbf{H}$ matrix, and back-transforming the output density matrix; however, it gains significantly with respect to the other methods in the initial SCF steps with many line searches per step.

\begin{table*}[t]
\begin{tabular*}{\textwidth}{lccccc}
\hline
\hline
\\[-10pt]
                      & No precon. & Precon. & Sparse, no precon. & Sparse, precon. & Cholesky \\
\\[-10pt]
\hline \\[-10pt]
$t/t^\mathrm{diagon}$ & 0.112      & 0.145   & 0.072              & 0.097           & 0.234    \\
\\[-10pt]
\hline
\hline
\end{tabular*}
\caption{Ratio of the average time taken for an SCF iteration with a single OMM line search $t$ and the average time taken for diagonalization $t^\mathrm{diagon}$, for the test case shown in Fig.~\ref{fig:single_step}.}
\label{table:single_step}
\end{table*}

For non-preconditioned OMMs, the large number of line searches needed in the first few SCF steps makes the method more expensive overall than diagonalization. When using preconditioning or Cholesky factorization, instead, it is faster by up to almost 60\% (considering only the time taken for the solver part of the DFT simulation).

Large, disordered systems such as liquid water generally present the greatest challenge for the OMM; crystalline systems with small unit cells and a fine Monkhorst-Pack (MP) k-point grid~\cite{mp_grid} are instead found to be the most favourable, since (a) small system sizes and crystalline order tend to require fewer line searches in the initial SCF steps, and (b) multiple k-points allow for a greater reuse of information, not only from one SCF step to the next, but also within a single SCF step from one k-point to another\footnote{However, the possibility of doing so is reduced when parallelizing the calculation across k-points; information reuse is maximized when all k-points are solved sequentially.}. Furthermore, our tests show a gain in relative efficiency for the OMM with respect to diagonalization when using complex matrices instead of real ones.

\begin{figure}[t]
\begin{center}\includegraphics{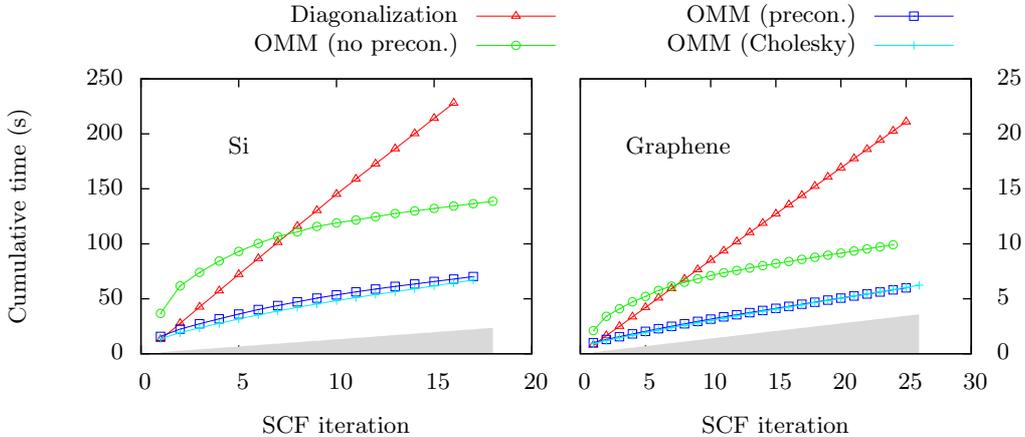}\end{center}
\caption{Timing tests for single-point energy calculations of two crystalline systems with k-point sampling. $\tau=10$~Ry for the preconditioned simulations.}
\label{fig:k}
\end{figure}

Fig.~\ref{fig:k} shows timing tests performed for two bulk crystalline systems: the 8-atom unit cell of silicon ($m=104$, $n=16$), and the 2-atom unit cell of graphene ($m=26$, $n=4$). We use MP grids of $9 \times 9 \times 9$ for silicon, and $19 \times 19 \times 1$ for graphene. The tests were run in serial. Results for the sparse routines are not included, as sparsity is negligible for such small systems. All the tested OMM flavours are faster than diagonalization for these two systems, with the preconditioned and Cholesky-factorized algorithms achieving speedups of 77\% and 78\%, respectively, for silicon, and 86\% and 85\% for graphene. The non-preconditioned algorithm now also results in an appreciable speedup, of 44\% for silicon and 63\% for graphene.

We note that the number of SCF iterations needed to reach the same convergence tolerance differs slightly between solvers; this can also be seen for the examples in Fig.~\ref{fig:precon}. This is due to small differences in the density matrix obtained at the end of each SCF step, which can be effectively eliminated by further reducing the OMM convergence threshold $e^\mathrm{tol}$ defined in Sec.~\ref{subsec:formal-conv-SCF}. Nevertheless, precoditionining/Cholesky factorization is observed to decrease the number of SCF iterations needed for a given $e^\mathrm{tol}$. We have also investigated the possibility of starting the simulation with a fairly high value of $e^\mathrm{tol}$, and progressively reducing it during the self-consistency cycle; in general, however, tests have shown that any saving obtained for the initial SCF steps in reducing the number of line searches is then lost due to an increase in the total number of SCF iterations.

\subsection{Hard and soft scaling}
\label{subsec:timing-scaling}

We now examine the efficiency of the OMM when parallelizing the calculation with MPI-2. This is simply related to the scaling efficiency of the underlying PBLAS and ScaLAPACK operations (and the sparse--dense multiplication, when used), and, hence, should be comparable to that of diagonalization with ScaLAPACK. Although the efficiency might vary significantly depending on the hardware and the underlying BLAS and MPI implementations used, we expect the relative trends between solvers to be fairly consistent. Our tests were performed on a BullX cluster with dual-processor Intel Xeon E5420 nodes and InfiniBand interconnects.

Both for the OMM solvers and diagonalization we employ a 2D block-cyclic data distribution of the matrices, with the exception of the sparse OMM algorithms, which employ a 1D block-cyclic distribution for compatibility with SIESTA's sparse matrix representation. Diagonalization is performed with a divide-and-conquer routine (\path{pdsyevd}/\allowbreak\path{pzheevd}), using the same multi-step process described previously for the Quickstep (CP2K) code in Ref.~\cite{VandeVondele2005103}. For the OMM solver with Cholesky factorization, the factorization itself is performed with the \path{pdpotrf}/\allowbreak\path{pzpotrf} routine, and the subsequent reduction of the generalized eigenvalue problem to standard form with the {\tt pdsygst}/{\tt pzhegst} routine. Instead, the OMM solvers with preconditioning make use of the \path{pdgetrf}/\allowbreak\path{pzgetrf} routine for the factorization, and the \path{pdgetri}/\allowbreak\path{pzgetri} routine for the inversion (a general routine must be used in this case, as $\mathbf{P}$ is not necessarily positive definite; in serial, however, the {\tt dsytrf}/{\tt zhetrf} and {\tt dsytri}/{\tt zhetri} routines are available).

It has previously been noted that the scaling efficiency of ScaLAPACK factorization and diagonalization can noticeably suffer when parallelizing over large numbers of cores ($\gtrsim 10^3$)~\cite{Rayson2008,Blum20092175,Auckenthaler2011783}, and several libraries~\cite{Hernandez2005,Auckenthaler2011783,Petschow2013} are being developed that can already outperform it (notably, the ELPA~\cite{Auckenthaler2011783} library within the FHI-aims~\cite{Blum20092175} and, recently, VASP~\cite{Kresse1996a} and CP2K~\cite{VandeVondele2005103} codes). Nevertheless, ScaLAPACK remains in wide usage, and is well established for benchmark tests.

\begin{figure}[t]
\begin{center}\includegraphics{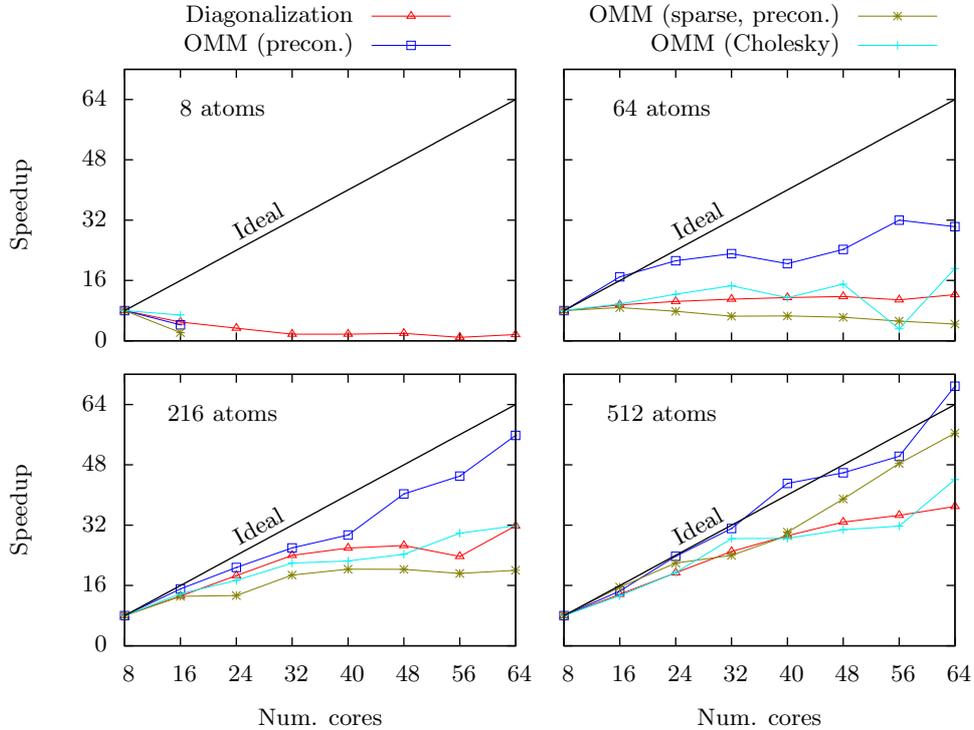}\end{center}
\caption{Hard scaling timing tests for bulk silicon. The speedup is given relative to the time taken on 8 cores for each solver. For the 8-atom system, only diagonalization can be used for $>16$ cores. $\tau=10$~Ry for the preconditioned simulations.}
\label{fig:hard}
\end{figure}

Firstly, we consider the case of hard scaling (Fig.~\ref{fig:hard}), i.e., increasing the number of cores while keeping the system size fixed. We do so for supercells of bulk silicon of 8, 64, 216, and 512 atoms, varying the number of cores from 8 to 64. The matrix dimensions are $m=13 N_a$ and $n=2 N_a$, where $N_a$ is the number of atoms. Fig.~\ref{fig:hard} shows the speedup relative to the 8-core timings, which we define as $8 t_8/t_{N_c}$, where $t_{N_c}$ is the total time spent by the solver for the self-consistent calculation on $N_c$ cores.

There is a general increase in scaling efficiency with system size, as this is mainly determined by the number of atoms/core; almost nothing can be gained with most solvers when decreasing this number to $\lesssim 10$. However, one OMM flavour (entirely dense algebra with preconditioning) is noticeably more efficient than both diagonalization and the other OMMs for all system sizes $>8$ atoms, giving timing gains even when going down to only a few atoms/core. For the 512-atom system, this method exhibits seemingly perfect hard scaling in the range of cores considered; in contrast, the efficiency of diagonalization is reduced to 55\% on 64 cores.

\begin{figure}[t]
\begin{center}\includegraphics{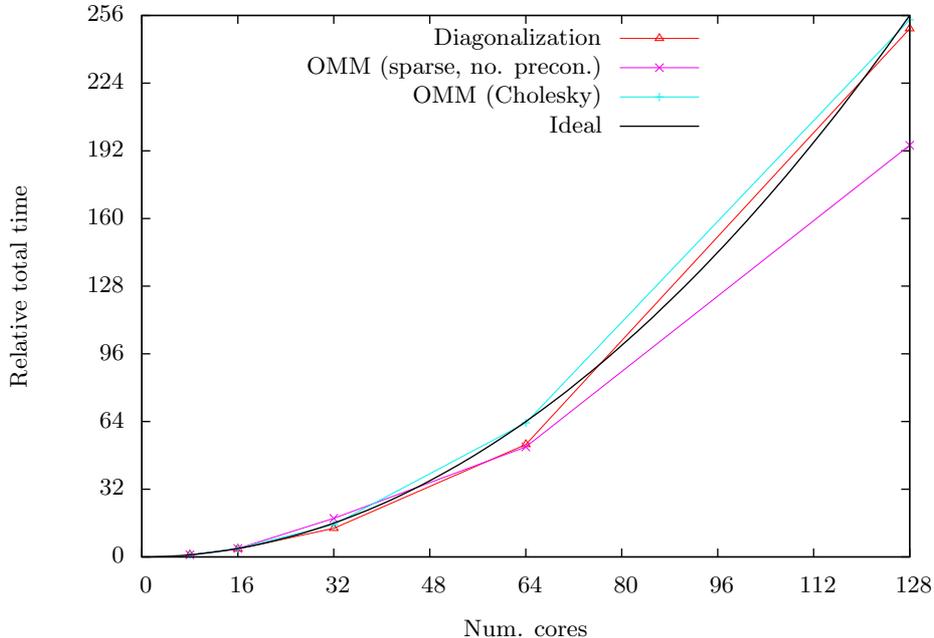}\end{center}
\caption{Soft scaling timing test for liquid water. The system size is proportional to the number of cores, with 8 molecules/core. The total time is given relative to that taken on 8 cores for each solver.}
\label{fig:soft}
\end{figure}

Of the two other OMM flavours considered, Cholesky factorization gives efficiencies similar to diagonalization, while the sparse routine with preconditioning is the least efficient for small systems, but shows a substantial increase in efficiency for the largest one. This is due to the level of sparsity of the $m \times m$ matrices increasing together with the system size. It is important to note that the speedup in Fig.~\ref{fig:hard} is only defined relative to the performance of the {\em same} solver on 8 cores; however, the OMMs are about twice as fast as diagonalization on this number of cores, meaning that all OMM flavours are actually consistently faster than diagonalization on all number of cores, even for cases in which the hard scaling efficiency is lower.

Next, we consider the case of soft scaling (Fig.~\ref{fig:soft}), i.e., keeping the number of atoms/core constant and increasing the system size together with the number of cores. For linear-scaling DFT codes, the aim in this case is to achieve a constant time-to-solution. Conventional cubic-scaling solvers, however, can at best achieve quadratic scaling, while an approximately constant time-to-solution requires a combination of hard and soft scaling.

For our test, we use snapshots of liquid water in simulations boxes of increasing size, with 8 molecules/core (within the range of efficient hard scaling suggested by the previous test) up to 128 cores. The matrix dimensions are $m=23 N_m$ and $n=4 N_m$, where $N_m$ is the number of molecules. Fig.~\ref{fig:soft} shows the timings relative to the 8-core (64-molecule) ones $t_{N_c}/t_8$, for each of three solvers. The ideal increase in time is therefore given by $N_c^2/64$, which is also shown in the figure. Both diagonalization and the OMM routine using entirely dense operations give timings that agree very closely with this ideal scaling, i.e., both methods exhibit essentially perfect soft scaling in the range of cores considered. The OMM routine with sparse--dense operations, instead, gives {\em better} than ideal scaling for the largest system; as for the hard scaling example, this is due to the increased sparsity of the $m \times m$ matrices, which reduces the percentage of the total time taken up by the \bbx\rct$=$\rct operations with respect to the smaller system sizes.

Both the hard and soft scaling tests, therefore, show the OMM to be at least as efficient as diagonalization, and potentially more so, depending on the specific OMM flavour. The improved parallel scaling of iterative solvers with respect to explicit diagonalization with ScaLAPACK has also been demonstrated with the AIMPRO~\cite{Rayson2008} and CP2K~\cite{VandeVondele2005103} codes. As we have already noted, however, newer libraries developed for massively parallel architectures are increasing the competitiveness of diagonalization and the range of system sizes for which it is feasible. Although some OMM algorithms are also dependent on ScaLAPACK operations that will undoubtably benefit from these same developments, the majority of the computational effort is concentrated on the matrix--matrix multiplications listed in Table~\ref{table:ops}, currently performed with the PBLAS {\tt pdgemm}/{\tt pzgemm} routine and our custom sparse--dense routine. Based on the results of our tests, the sparse version of the OMM solver appears to be the most promising for massive parallelization; this is also the strategy pursued by CP2K for a similar iterative solver and localized basis~\cite{VandeVondele2005103}. To this end, we can identify certain key improvements that need to be made to our current implementation: (a) a sparsity-preserving factorization (and, hence, a sparse preconditioner), and (b) a new sparse matrix representation for SIESTA, compatible with the 2D block-cyclic data distribution scheme used by PBLAS, which exhibits better scaling to large numbers of cores than the 1D one.

\subsection{Basis size scaling}
\label{subsec:timing-basis}

We conclude our timing tests by considering the scaling of the OMM with the basis size $m$, while keeping the system size (and the number of occupied eigenstates $n$) constant. For this test, we move away from pure atomic basis sets, and use instead a hybrid basis, consisting of the usual atom-centred orbitals, plus sets of spherical Bessel functions confined to overlapping spheres and fixed in space to a regular grid, with each sphere centred on a grid point. In doing so, we can reach very large basis sizes (approaching those of plane-wave calculations), and demonstrate the qualitatively different scaling behaviour of diagonalization, entirely dense and sparse--dense OMMs, this last method being optimally suited to take advantage of the properties of high-quality, localized and variational basis sets such as blips~\cite{Hernandez1997} or psincs~\cite{Mostofi2002}.

\begin{figure}[t]
\begin{center}\includegraphics{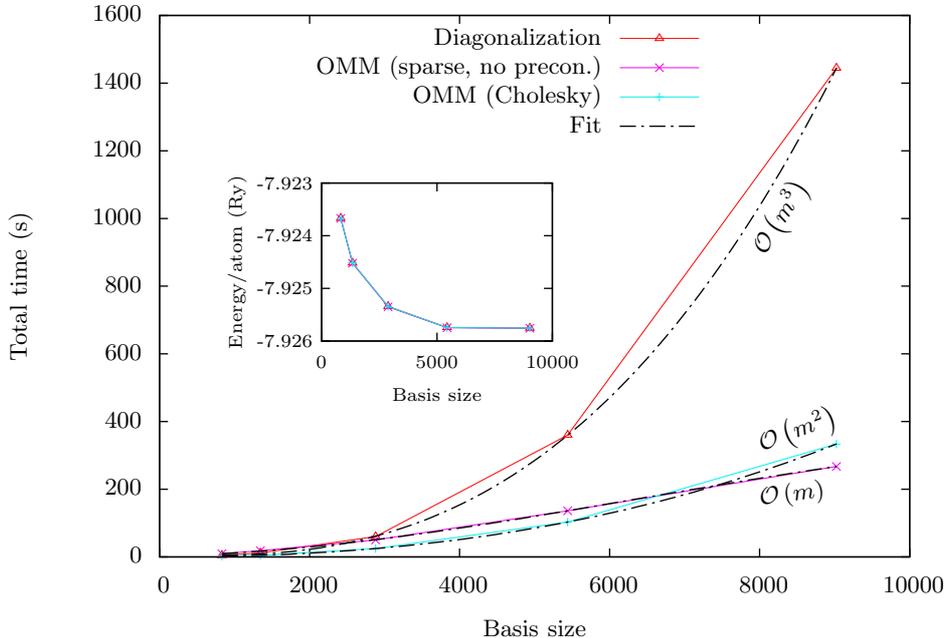}\end{center}
\caption{Basis size scaling timing test for 64 atoms of bulk silicon using a hybrid basis. The black dash--dotted lines show the fits to the data points for the three different solvers; a third-order polynomial is used in each case.}
\label{fig:bessel}
\end{figure}

Fig.~\ref{fig:bessel} shows the total time taken by three different solvers for a self-consistent calculation of 64 atoms of bulk silicon, performed on 16 cores. We use the standard \DZP\ basis for the atomic orbitals, and Bessel functions fixed to an $8 \times 8 \times 8$ grid, offset by half a grid spacing. The Bessel functions are confined within spheres of radius $r_B=g\sqrt{3}/2$, where $g$ is the grid spacing. The basis size is increased by adding shells of Bessel functions of higher $l$, up to $l=3$. The fraction of occupied eigenstates varies from 15\% down to 1.4\%.

The polynomial fits in the figure show that the three solvers follow different scaling behaviour with basis size: diagonalization is cubic, the OMM with entirely dense algebra is quadratic, and the OMM with sparse--dense algebra is linear. This is straightforward to explain from the list of operations in Table~\ref{table:ops}: only \bbx\rct$=$\rct scales as $n m^2$, while all other operations scale at worst as $n^2 m$; when using sparse algebra, however, the former is reduced to $n m$.

Of course, this scaling behaviour is not unique to the OMM, but is shared with all other iterative minimization methods. Achieving linear scaling up to large basis sizes, however, requires not only a localized basis, but also a sparse preconditioner, or (as in this example) a method that is sufficiently efficient even without preconditioning. This statement might seem surprising, given the examples discussed in Sec.~\ref{subsec:formal-conv-opt}; however, we have found that, while the ill-conditioning is very severe when increasing the number of shells of the atomics orbitals, it is much less affected by the grid-based orbitals. In fact, in the example shown in Fig.~\ref{fig:bessel}, the number of line searches needed for convergence increases only by a factor of $\sim$2 between the smallest basis ($m=832$) and the largest one ($m=9024$), while the width of the eigenspectrum increases by a factor of $\sim$8. By comparison, for the system shown in Fig.~\ref{fig:eta}, there is a factor of $\sim$300 increase in the number of line searches for a similar increase in the eigenspectrum width.

\section{Conclusions}
\label{sec:outro}

We have presented a cubic-scaling implementation of the OMM for finite-range atomic basis sets within a self-consistent KS solver. Particular attention has been given to the number and type of matrix operations needed for each line search step, and a number of different variants of the main algorithm have been proposed, optionally making use of the sparsity of the Hamiltonian and overlap matrices to reduce the computational cost while retaining the full accuracy of the solution. The use of a preconditioning scheme for localized orbitals has been investigated, and has been found to be effective in stabilizing the number of iterations needed for convergence between different basis sizes and system types.

Timing tests for self-consistent calculations have shown the OMM to be able to achieve a greater efficiency than explicit diagonalization even for minimal basis sizes, with the greatest speedups (up to almost 90\%) being found for small periodic systems with multiple k-points; this is mainly due to the possibility of information reuse between subsequent minimizations within the self-consistency cycle, typically resulting in a single line search being needed for the last few SCF iterations. Information reuse can also be employed between different single-point energy calculations, e.g. for MD simulations or geometry optimizations. Future work on our implementation of the method in SIESTA will focus on supporting finite-temperature (Fermi level smearing) calculations.

\section*{Acknowledgments}

We acknowledge useful discussions with Emilio Artacho, Jos\'{e} M. Soler, Alberto Garc\'{i}a, Pablo Ordej\'{o}n, Georg Huhs, Rogeli Grima, Jos\'{e} M. Cela, and Peter D. Haynes. The calculations were performed on the following HPC clusters: kroketa, tortilla (CIC nanoGUNE, Spain), arina (Universidad del Pa\'{i}s Vasco/Euskal Herriko Unibertsitatea, Spain), altamira (Universidad de Cantabria, Spain). We thank the RES--Red Espa\~nola de Supercomputaci\'{o}n for access to altamira. SGIker (UPV/EHU, MICINN, GV/EJ, ERDF and ESF) support is gratefully acknowledged.

\section*{References}

\bibliographystyle{elsarticle-num.bst}

\end{document}